# The easiest way to Heaviside ellipsoid

## Valery P. Dmitriyev

*Lomonosov University*
*P.O.Box 160, Moscow 117574, Russia*
*e-mail: dmitr@cc.nifhi.ac.ru*

**Abstract**. The formula for the electric field of a point charge moving with constant velocity is derived using the symmetry properties of Maxwell's equations - its Lorentz invariance. In contrast to conventional treatments, the derivation presented does not use retarded integrals or relativity transformations.

We are interested in a simple and concise derivation of the formula for the electromagnetic field produced by an electric charge moving with constant velocity. The standard textbook approach is commonly based on the relativistic transformation of the fields. The whole case looks such as if classical electrodynamics is incomplete and needs external facilities in order to derive some of its formulae. Really, of course, electrodymanics is a consistent theory and all necessary relations can be obtained from Maxwell's equations without recourse to any extraneous postulates. Recently Prof. Jefimenko has demonstrated that in a series of works. However, below the method is proposed which is more straightforward and matter-of-course comparing with what was given in [1].

Nowadays the symmetry properties of physical systems are usually brought to the forefront. Frequently, they receive the rights of their own and lay the basis for the whole theory, as it was first with special relativity. However, sometimes, when needed, they don't find proper implementation. In this connection I would like to emphasize another time that the primary destination of symmetry relationships is to make easier the procedure of integrating the equations.

We proceed from the wave equations for electromagnetic potentials **A** and $\varphi$

$$\nabla^2 \mathbf{A} - \frac{1}{c^2}\frac{\partial^2 \mathbf{A}}{\partial t^2} = -\frac{4\pi}{c}\mathbf{j} \qquad (1)$$

$$\nabla^2 \varphi - \frac{1}{c^2}\frac{\partial^2 \varphi}{\partial t^2} = -4\pi\rho \qquad (2)$$

They are easily obtained from Maxwell's equations

$$\mathbf{E} = -\nabla\varphi - \frac{1}{c}\frac{\partial \mathbf{A}}{\partial t} \qquad (3)$$

$$\frac{\partial \mathbf{E}}{\partial t} - c\nabla\times(\nabla\times\mathbf{A}) + 4\pi\mathbf{j} = 0$$

$$\nabla \cdot \mathbf{E} = 4\pi\rho$$

combining it with the Lorentz gauge

$$\nabla \cdot \mathbf{A} + \frac{1}{c}\frac{\partial \varphi}{\partial t} = 0$$

Now, we have to solve equations (1), (2) and then using (3) get the necessary formulae. For the source, moving with the velocity $\mathbf{v} = \text{const}$, the charge density is

$$\rho(\mathbf{x} - \mathbf{v}t)$$

and the current density

$$\mathbf{j} = \mathbf{v}\rho(\mathbf{x} - \mathbf{v}t)$$

It implies that the electromagnetic potentials are also the functions of $\mathbf{x} - \mathbf{v}t$. With this one may take advantage of the symmetry properties of the system (1), (2).

The basic fact is that the left-hand side of the inhomogeneous d'Alambert equation

$$\nabla^2 f - \frac{1}{c^2}\frac{\partial^2 f}{\partial t^2} = g(\mathbf{x} - \mathbf{v}t) \qquad (4)$$

is Lorentz-invariant. That enables us to reduce the kinetic problem to the static one. However, in order to do this, one does not need the whole Lorentz transformation. It suffices to use the part of it:

$$x_1' = \frac{x_1 - vt}{\sqrt{1 - v^2/c^2}} \qquad (5)$$

$$t' = t$$

Passing in (4) to the reference frame (5) gives:

$$\frac{\partial f}{\partial x} = \frac{\partial f}{\partial x'} \cdot \frac{\partial x'}{\partial x} = \frac{\partial f}{\partial x'}\gamma$$

$$\frac{\partial^2 f}{\partial x^2} = \frac{\partial^2 f}{\partial x'^2}\gamma^2$$

$$\frac{\partial f}{\partial t} = \frac{\partial f}{\partial t'} + \frac{\partial f}{\partial x'}\cdot\frac{\partial x'}{\partial t} = \frac{\partial f}{\partial t'} - \frac{\partial f}{\partial x'}v\gamma$$

$$\frac{\partial^2 f}{\partial t^2} = \frac{\partial^2 f}{\partial t'^2} + \frac{\partial^2 f}{\partial x'\partial t}\cdot\frac{\partial x'}{\partial t} - \frac{\partial^2 f}{\partial t'\partial x'}v\gamma - \frac{\partial^2 f}{\partial x'^2}\cdot\frac{\partial x'}{\partial t}v\gamma$$

$$= \frac{\partial^2 f}{\partial t'^2} - 2v\gamma\frac{\partial^2 f}{\partial t'\partial x'} + v^2\gamma^2\frac{\partial^2 f}{\partial x'^2}$$

where

$$\gamma = (1 - v^2/c^2)^{-1/2}$$

Hence

$$\frac{\partial^2 f}{\partial x^2} - \frac{1}{c^2}\frac{\partial^2 f}{\partial t^2} = \frac{\partial^2 f}{\partial x'^2}\gamma^2 - \frac{v^2\gamma^2}{c^2}\frac{\partial^2 f}{\partial x'^2} + \frac{2v\gamma}{c^2}\frac{\partial^2 f}{\partial x'\partial t'} - \frac{1}{c^2}\frac{\partial^2 f}{\partial t'^2} = \frac{\partial^2 f}{\partial x'^2} + \frac{2v\gamma}{c^2}\frac{\partial^2 f}{\partial x'\partial t'} - \frac{1}{c^2}\frac{\partial^2 f}{\partial t'^2}$$

Taking into account that in new frames

$$\frac{\partial f}{\partial t'} = 0$$

we get from (4) the Poisson equation

$$\frac{\partial^2 f}{\partial x_1'^2} + \frac{\partial^2 f}{\partial x_2} + \frac{\partial^2 f}{\partial x_3} = g(x_1'\gamma^{-1}, x_2, x_3)$$

Following this line let us consider the motion of a point electric charge. In this event the set of the equations (1), (2) looks as

$$\nabla^2 \mathbf{A} - \frac{1}{c^2}\frac{\partial^2 \mathbf{A}}{\partial t^2} = -\frac{4\pi \mathbf{v}}{c} q\delta(\mathbf{x} - \mathbf{v}t)$$

$$\nabla^2 \varphi - \frac{1}{c^2}\frac{\partial^2 \varphi}{\partial t^2} = -4\pi q\delta(\mathbf{x} - \mathbf{v}t)$$

Passing to the reference frame (5), which moves uniformly along the axis $x_1$ together with the charge, we get

$$\frac{\partial^2 A_1}{\partial x_1'^2} + \frac{\partial^2 A_1}{\partial x_2} + \frac{\partial^2 A_1}{\partial x_3} = -4\pi q\gamma \frac{\mathbf{v}}{c}\delta(x_1', x_2, x_3) \quad (6)$$

$$\frac{\partial^2 \varphi}{\partial x_1'^2} + \frac{\partial^2 \varphi}{\partial x_2} + \frac{\partial^2 \varphi}{\partial x_3} = -4\pi q\gamma\delta(x_1', x_2, x_3) \quad (7)$$

In the right-hand sides of (6) and (7) the following property of the $\delta$ - function was used

$$\delta(ax) = \frac{1}{a}\delta(x), \quad a > 0$$

Using the relation

$$\nabla^2 \frac{1}{|\mathbf{x}|} = -4\pi\delta(\mathbf{x})$$

the static problem (6), (7) is easily resolved:

$$A_1 = \frac{v}{c} \cdot \frac{q\gamma}{R}, \quad A_2 = 0, \quad A_3 = 0 \quad (8)$$

$$\varphi = \frac{q\gamma}{R} \quad (9)$$

where

$$R = \left[\gamma^2(x_1 - vt)^2 + x_2^2 + x_3^2\right]^{1/2} \quad (10)$$

Next, we calculate the portions for (3)

$$\frac{\mathbf{i_1}}{c} \cdot \frac{\partial A_1}{\partial t} = \mathbf{i}_1 q\gamma \cdot \frac{v^2}{c^2} \cdot \frac{\gamma^2(x_1 - vt)}{R^3}$$

$$\nabla\varphi = -q\gamma \cdot \frac{\mathbf{i}_1 \gamma^2 (x_1 - vt) + \mathbf{i}_2 x_2 + \mathbf{i}_3 x_3}{R^3}$$

whence

$$\mathbf{E} = q\gamma \cdot \frac{\mathbf{i}_1 \gamma^2 (x_1 - vt)(1 - v^2/c^2) + \mathbf{i}_2 x_2 + \mathbf{i}_3 x_3}{R^3}$$

$$= q\gamma \cdot \frac{\mathbf{i}_1 (x_1 - vt) + \mathbf{i}_2 x_2 + \mathbf{i}_3 x_3}{R^3} \tag{11}$$

In spherical coordinates we have

$$x_1 - vt = r\cos\theta, \qquad x_2^2 + x_3^2 = r^2 \sin\theta$$

where $\theta$ is the angle between the radius vector $\mathbf{r} = \mathbf{i}_1(x_1 - vt) + \mathbf{i}_2 x_2 + \mathbf{i}_3 x_3$ and $x_1$ axis. Thus

$$R^2 = \gamma^2(x_1 - vt)^2 + x_2^2 + x_3^2 = \gamma^2 r^2 \cos^2\theta + r^2 \sin^2\theta = \gamma^2 r^2 \left(1 - \frac{v^2}{c^2}\sin^2\theta\right)$$

Using it in (11) we find finally

$$E = \frac{q\left(1 - \dfrac{v^2}{c^2}\right)}{r^2 \left(1 - \dfrac{v^2}{c^2}\sin^2\theta\right)^{3/2}} \tag{12}$$

The latter is just the famous Heaviside formula. It describes the real physical effect of the "squashing" the electric field against the direction of motion:

$$E(\theta = 0) = \frac{q}{r^2} \cdot (1 - v^2/c^2)$$

$$E\left(\theta = \frac{\pi}{2}\right) = \frac{q}{r^2} \cdot \frac{1}{(1 - v^2/c^2)^{1/2}}$$

So that

$$\frac{E(0)}{E\left(\dfrac{\pi}{2}\right)} = \left(1 - \frac{v^2}{c^2}\right)^{3/2}$$

I wonder why the derivation of (12) presented did not become common for textbooks.

At last, let us find the total electromagnetic force field generated by the moving charge $q$. We have from (8), (9)

$$\mathbf{A} = \frac{\mathbf{v}}{c} \cdot \varphi$$

Hence

$$\mathbf{H} = \nabla \times \mathbf{A} = \frac{1}{c} \cdot \nabla \times (\mathbf{v}\varphi) = \frac{1}{c} \cdot (\nabla \varphi \times \mathbf{v})$$

From (3)

$$\mathbf{H} = -\frac{1}{c} \cdot \left(\mathbf{E} + \frac{1}{c}\frac{\partial \mathbf{A}}{\partial t}\right) \times \mathbf{v} = -\frac{1}{c}\mathbf{E} \times \mathbf{v} - \frac{1}{c^2}\frac{\partial \varphi}{\partial t}\mathbf{v} \times \mathbf{v} = \frac{1}{c} \cdot \mathbf{v} \times \mathbf{E}$$

The total force on a charge $q_0$ is given by

$$\mathbf{F} = q_0\left[\mathbf{E} + \frac{1}{c}(\mathbf{v} \times \mathbf{H})\right] = q_0\left[\mathbf{E} - \frac{v^2}{c^2}\mathbf{E} + \frac{\mathbf{v}(\mathbf{v}\mathbf{E})}{c^2}\right]$$

$$= q_0 q \frac{\mathbf{i}_1 \gamma(x_1 - vt) + \mathbf{i}_2 x_2 + \mathbf{i}_3 x_3}{\left[\gamma^2(x_1 - vt)^2 + x_2^2 + x_3^2\right]^{3/2}} = -q_0 q \nabla' \frac{1}{R} = -q_0 \nabla' \psi$$

That is just the formula for the Heaviside ellipsoid $\psi = \text{const}$, where $\psi = q/R$, $R$ is given by (10) and gradient $\nabla'$ is taken in moving coordinates (5). So, you see that the total electromagnetic force field is undergone the real physical effect of the Lorentz contraction along the direction of motion.